\documentclass[onecolumn,showpacs,preprintnumbers]{revtex4}
\usepackage{graphicx}
\usepackage{dcolumn}
\usepackage{bm}
\usepackage{epsfig}
\usepackage{amsmath,amssymb}

\begin{document}
\title{Analysis of the charmed mesons $D_{1}^{*}(2680)$, $D_{3}^{*}(2760)$ and $D_{2}^{*}(3000)$
}
\author{Guo-Liang Yu$^{1}$}
\email{yuguoliang2011@163.com}
\author{Zhi-Gang Wang$^{1}$}
\email{zgwang@aliyun.com}
\author{ZhenYu Li$^{2}$}

\affiliation{$^1$ Department of Mathematics and Physics, North China
Electric power university, Baoding 071003, People's Republic of China\\$^2$
School of Physics and Electronic Science, Guizhou Normal College,
Guiyang 550018, People's Republic of China}
\date{\today }

\begin{abstract}
In this work, we systematically study the strong decay behavior of
the charmed mesons $D_{1}^{*}(2680)$, $D_{3}^{*}(2760)$ and
$D_{2}^{*}(3000)$ reported by the LHCb collaboration. By comparing
the masses and the decay properties with the results of the
experiment, we assign these newly observed mesons as the
$2S\frac{1}{2}1^{-}$, $1D\frac{5}{2}3^{-}$ and $1F\frac{5}{2}2^{+}$
states respectively. As a byproduct, we also study the strong decays
of the unobserved $2P\frac{3}{2}2^{+}$, $2F\frac{5}{2}2^{+}$ and
$3P\frac{3}{2}2^{+}$ charmed mesons, which is useful for future
experiments in searching for these charmed mesons.
\end{abstract}


\maketitle

\begin{large}
\textbf{1 Introduction}
\end{large}

Recently, the LHCb Collaboration studied the resonant substructures
of $B^{-}$$\rightarrow$$D^{+}$$\pi^{-}$$\pi^{-}$ decays in a data
sample corresponding to $3.0$ $fb^{-1}$ of pp collision data
recorded by the LHCb experiment during 2011 and 2012. By a Dalitz
plot analysis technique, the presence of resonances with spins 1, 2
and 3 at the $D^{+}$$\pi^{-}$ mass spectrum were
confirmed~\cite{Aaij}. Their analysis indicated that these
resonances are mainly from the contributions of $D_{2}^{*}(2460)$,
$D_{1}^{*}(2680)$, $D_{3}^{*}(2760)$ and $D_{2}^{*}(3000)$ charmed
mesons. The masses and decay widths of these  mesons are

$D_{2}^{*}(2460):M=2463.7\pm0.4\pm0.4\pm0.6$MeV, $\Gamma=47.0\pm0.8\pm0.9\pm0.3$MeV

$D_{1}^{*}(2680):M=2681.1\pm5.6\pm4.9\pm13.1$MeV, $\Gamma=186.7\pm8.5\pm8.6\pm8.2$MeV

$D_{3}^{*}(2760):M=2775.5\pm4.5\pm4.5\pm4.7$MeV, $\Gamma=95.3\pm9.6\pm7.9\pm33.1$MeV

$D_{2}^{*}(3000):M=3214\pm29\pm33\pm36$MeV, $\Gamma=186\pm38\pm34\pm63$MeV

Actually, people have found many other charmed mesons before these
discoveries~\cite{N1,N2,N3,N4,N5,N6,N7,N8} , which have greatly
enriched the charmonium spectra. On the other hand, these
discoveries also shed more light on our knowledge about the essence
of the elementary particles in the micro-world. For
$D_{2}^{*}(2460)$ as an example, it has been well established
previously and the $1P\frac{3}{2}2^{+}$ assignment is strongly
favored~\cite{N9}. We studied the nature of the states
$D_{1}^{*}(2680)$, $D_{3}^{*}(2760)$ and $D_{2}^{*}(3000)$ in our
previous work using the heavy meson effective theory~\cite{WZG1}.
Some of the strong decay behavior have also been studied in which
the calculated ratios among the decay widths can be used to confirm
or reject the assignments of the newly observed charmed mesons. The
decay behavior of the $D_{2}^{*}(3000)$ charmed meson was also
analyzed in reference~\cite{JZW}, where it was assigned as the
2$^{3}F_{2}$ or 3$^{3}P_{2}$ states. In order to identify the
$D_{1}^{*}(2680)$, $D_{3}^{*}(2760)$ and $D_{2}^{*}(3000)$ and give
more specific decay widths and the ratios, we further analyze the
strong decay properties of these newly observed charmed mesons using
the $^{3}P_{0}$ decay model.

The $^{3}P_{0}$ decay model is known as quark pair creation model
(QPC) which was firstly introduced by Micu~\cite{Micu} in 1969. An
important feature of the this decay model, apart from its
simplicity, is that it provides the gross features of several
transitions with two parameters, the pair-crestion strength $\gamma$
and the oscillator parameter $R$, which can be fitted to the
experimental data. Soon after the introduction of the $^{3}P_{0}$
model, it was further developed by other
collaborations~\cite{Car,Le}. This model, extensively applied to the
decays of light mesons and
baryons~\cite{Yao77,Yaou88,Robe92,Capstick,Blun,Zhou05,Chen09,Dml08,Bzh07,Yang10},
has been applied to evaluate the strong decays of heavy meson in the
charmonium~\cite{Barn,Fer1,Fer2}, bottomonium~\cite{Fer2,Fer3}, and
open-charm sectors~\cite{Geiger,Sego}.

Just as what we have analyzed~\cite{WZG1}, the mesons of
$D_{1}^{*}(2680)$, $D^{*}(2600)$ and $D_{J}^{*}(2650)$ have the
similar mass and width~\cite{M1,M2}, and can be assigned to be the
same states $2S\frac{1}{2}1^{-}$~\cite{M3,M4,M5,M6,M7}. Based on the
same analysis, $D_{3}^{*}(2760)^{0}$, $D^{*}(2760)^{0}$,
$D_{J}^{*}(2760)^{0}$ may be the same particle, and can be assigned
to be the $1D\frac{5}{2}3^{-}$ state
~\cite{M1,M2,M3,M4,M5,M6,M7,M8,M9}. As for $D_{2}^{*}(3000)$, it can
be a $P$ wave and $F$ wave charmed meson. Its mass can be calculated
by different theoretical models, such as the relativized quark model
based on a universal one-gluon exchange plus linear confinement
potential~\cite{Godf}, the relativistic quark model includes the
leading order $1/M_{h}$ corrections ~\cite{Pierro}, the
QCD-motivated relativistic quark model based on the quasipotential
approach~\cite{Eber}. According to these calculations,
$1F\frac{5}{2}2^{+}$, $2P\frac{3}{2}2^{+}$, $2F\frac{5}{2}2^{+}$ and
$3P\frac{3}{2}2^{+}$ can also be assigned as the candidates of the
possible states of the charmed meson $D_{2}^{*}(3000)$.

To further verify the states of $D_{1}^{*}(2680)$ and
$D_{3}^{*}(2760)$ and check the possibilities of different
assignments of the $D_{2}^{*}(3000)$, we give a systematic analysis
of the decay behaviors about these charmed mesons. The article is
arranged as follows: In section 2, the brief review of the
$^{3}P_{0}$ decay model is given (For the detailed review see
Refs.~\cite{Le,Yaou88,Robe92,Blun}); in Sec.3, we study the strong
decays of the charmed mesons $D_{1}^{*}(2680)$, $D_{3}^{*}(2760)$
and $D_{2}^{*}(3000)$ observed by the LHCb collaboration with the
$^{3}P_{0}$ decay model; in Sec.4, we present our conclusions.

\begin{large}
\textbf{2 METHOD}
\end{large}

\begin{large}
\textbf{2.1 The decay model}
\end{large}

\begin{figure}[h]
  \includegraphics[width=15cm]{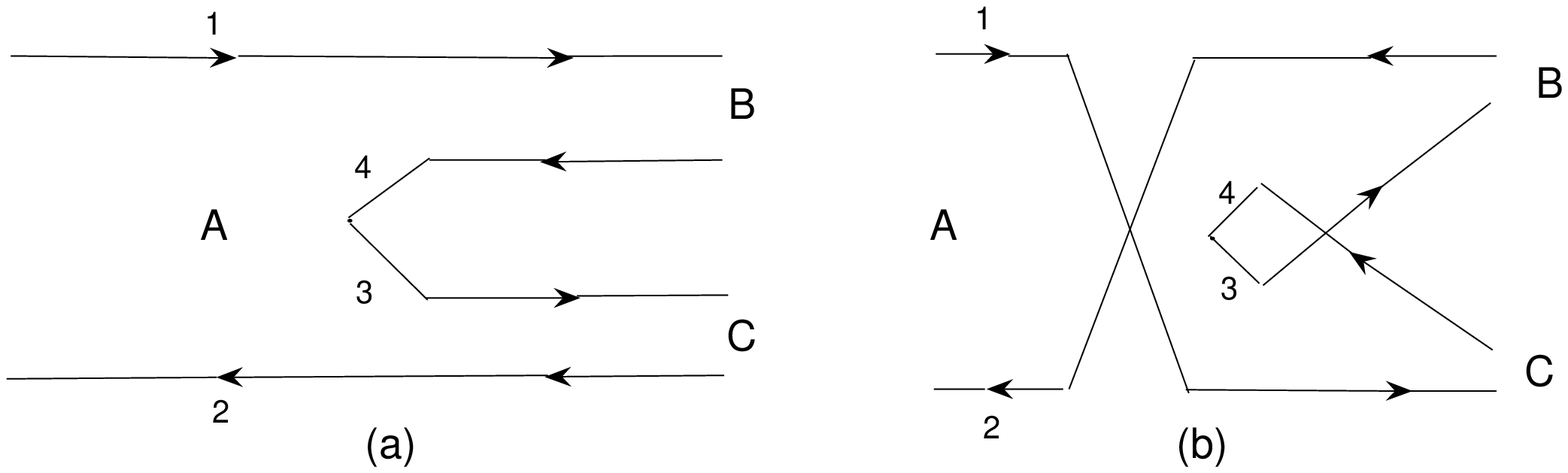}
  \caption{The two possible decay processes of $A\rightarrow BC$ in the $^{3}P_{0}$ model.}\label{Figure 1:}
\end{figure}

The $^{3}P_{0}$ decay model assumes that a quark-antiquark pair is
created from the vacuum with the corresponding quantum number
$0^{++}$. This new $q\overline{q}$ together with the $q\overline{q}$
within the initial meson regroups into two outgoing mesons in all
possible arrangements for the meson decay process
$A$$\rightarrow$$BC$ as shown in Fig. 1.

In the nonrelativistic limit, the transition operator of this
process can be expressed as
\begin{equation}
\begin{split}
    T=
    &-3\gamma\sum_{m}\langle1m1-m\mid00\rangle\int d^{3}\vec p_{3}d^{3}\vec p_{4}\delta^{3}(\vec p_{3}+\vec p_{4})\mathcal{Y}_{1}^{m}(\frac{\vec p_{3}-\vec p_{4}}{2})
    \chi_{1-m}^{34}\varphi_{0}^{34}\omega_{0}^{34}b_{3}^{\dag}(\vec p_{3})d_{4}^{\dag}(\vec p_{4})
\end{split}
\end{equation}
where the dimensionless parameter $\gamma$ denotes the creation
strength of the quark-antiquark $q_{3}\overline{q}_{4}$ pair. $\vec
p_{3}$ and $\vec p_{4}$ are the momenta of this quark-antiquark
pair. Its flavor, color, and spin wave functions are represented by
$\varphi_{0}^{34}$, $\omega_{0}^{34}$, and $\chi_{1-m}^{34}$,
respectively. $\mathcal{Y}_{1}^{m}(\vec p)\equiv|\vec
p|^{1}Y_{1}^{m}(\theta_{p},\phi_{p})$ is a solid harmonic polynomial
corresponding to the p-wave quark pair.

In the center of mass frame of parent meson $A$, the helicity
amplitude $\mathcal{M}^{M_{J_{A}}M_{J_{B}}M_{J_{C}}}$ of the decay
process $A\rightarrow BC$ is written as
\begin{equation}
\begin{aligned}
\mathcal{M}^{M_{J_{A}}M_{J_{B}}M_{J_{C}}}(\vec P)=
&\gamma\sqrt{8E_{A}E_{B}E_{C}}\sum_{\mbox{\tiny$\begin{array}{c}
M_{L_{A}},M_{S_{A}},\\
M_{L_{B}},M_{S_{B}},\\
M_{L_{C}},M_{S_{C}},m\end{array}$}}\langle
L_{A}M_{L_{A}}S_{A}M_{S_{A}}\mid J_{A}M_{J_{A}}\rangle \langle
L_{B}M_{L_{B}}S_{B}M_{S_{B}}\mid J_{B}M_{J_{B}}\rangle \\
&\times\langle L_{C}M_{L_{C}}S_{C}M_{S_{C}}\mid
J_{C}M_{J_{C}}\rangle\langle 1m1-m\mid 00\rangle\langle \chi_{S_{B}M_{S_{B}}}^{14}\chi_{S_{C}M_{S_{C}}}^{32}\mid \chi_{S_{A}M_{S_{A}}}^{12}\chi_{1-m}^{34}\rangle \\
&\times[\langle \phi_{B}^{14}\phi_{C}^{32}\mid \phi_{A}^{12}\phi_{0}^{34}\rangle I(\vec P,m_{1},m_{2},m_{3}) \\
&+(-1)^{1+S_{A}+S_{B}+S_{C}}\langle \phi_{B}^{32}\phi_{C}^{14}\mid \phi_{A}^{12}\phi_{0}^{34}\rangle I(-\vec P,m_{2},m_{1},m_{3})]
\end{aligned}
\end{equation}
where the spatial integral is defined as
\begin{equation}
\begin{split}
I(\vec P,m_{1},m_{2},m_{3})=
&\int d^{3}\vec p \psi^{*}_{n_{B}L_{B}M_{L_{B}}}(\frac{m_{3}}{m_{1}+m_{2}}\vec P_{B}+\vec p)\psi^{*}_{n_{C}L_{C}M_{L_{C}}}(\frac{m_{3}}{m_{2}+m_{3}}\vec P_{B}+\vec p) \\
&\times\psi_{n_{A}L_{A}M_{L_{A}}}(\vec P_{B}+\vec p)\mathcal{Y}_{1}^{m}(\vec p)
\end{split}
\end{equation}
where $\vec P =\vec P_{B} =-\vec P_{C}, \vec p = \vec p_{3}$, $m_{3}$ is the mass of the created quark $q_{3}$, the simple harmonic oscillator (SHO)
approximation is used for the meson space wave functions:
\begin{equation}
\begin{split}
\Psi_{nLM_{L}}(\vec p)=
&(-1)^{n}(-i)^{L}R^{L+\frac{3}{2}}\sqrt{\frac{2n!}{\Gamma(n+L+\frac{3}{2})}}exp(-\frac{R^{2}p^{2}}{2})L_{n}^{L+\frac{1}{2}}(R^{2}p^{2})\mathcal{Y}_{LM_{L}}(\vec
p)
\end{split}
\end{equation}

The partial wave amplitudes are related to the helicity amplitudes by~\cite{Jaco}
\begin{equation}
\begin{split}
\mathcal{M}^{JL}(\vec P)=
&\frac{\sqrt{4\pi(2L+1)}}{2J_{A}+1}\sum_{M_{J_{B}}M_{J_{C}}}\langle
L0JM_{J_{A}}|J_{A}M_{J_{A}}\rangle \langle
J_{B}M_{J_{B}}J_{C}M_{J_{C}}|JM_{J_{A}}\rangle\mathcal{M}^{M_{J_{A}}M_{J_{B}}M_{J_{C}}}(\vec
P)
\end{split}
\end{equation}
where $M_{J_{A}}=M_{J_{B}}+M_{J_{C}}$, $\mathbf{J_{A}=J_{B}+J_{C}}$
and $\mathbf{J_{A}+J_{P}=J_{B}+J_{C}+J_{L}}$. The transition in
terms of partial wave amplitudes is
\begin{equation}
\Gamma=\frac{\pi}{4}\frac{|\vec P|}{M_{A}^{2}}\sum_{JL}|\mathcal{M}^{JL}|^{2}
\end{equation}
where $P=|\vec
P|=\frac{\sqrt{[M_{A}^{2}-(M_{B}+M_{C})^{2}][M_{A}^{2}-(M_{B}-M_{C})^{2}]}}{2M_{A}}$,
$M_{A}$, $M_{B}$, and $M_{C}$ are the masses of the meson $A$, $B$,
and $C$.

\begin{large}
\textbf{2.2 Mixed states}
\end{large}

Heavy-light mesons are not charge conjugation eigenstates and so
mixing can occur among states with the same $J^{P}$ that are
forbidden for neutral states~\cite{ADe}. These occur between states
with $J=L$ and $S=1$ or $0$~\cite{ADe,Mats}. When $J=L=1$, the
corresponding mixture angle is $\theta=-54.7^{\circ}$ or
$\theta=35.3^{\circ}$~\cite{ADe,Mats}. The two $1^{+}$ charmed
mesons are the mixtures of the $^{3}P_{1}$ and $^{1}P_{1}$ states:
\begin{equation}
\begin{pmatrix} |\frac{1}{2},1^{+}\rangle &  \\ |\frac{3}{2},1^{+}\rangle  \end{pmatrix} \\=\begin{pmatrix} \cos\theta &  -\sin\theta  \\ \sin\theta & \cos\theta   \end{pmatrix} \\\begin{pmatrix} |^{3}P_{1}\rangle &  \\ |^{1}P_{1}\rangle  \end{pmatrix} \\
\end{equation}
In our calculation, the final states are related to
$D(2420)/D(2430)$ and $D_{s_{1}}(2460)/D_{s_{1}}(2536)$, which are
the $1^{+}$ states in the $D$ and $D_{s}$ meson families,
respectively. $D(2420)/D(2430)$ and
$D_{s_{1}}(2460)/D_{s_{1}}(2536)$ are the mixing of the $^{3}P_{1}$
and $^{1}P_{1}$ states, which satisfy the above relation(see Eq.7).
Thus the helicity amplitude can also be deduced as follows
\begin{equation}
\begin{pmatrix} \mathcal{M}^{JL}_{|A\rangle\rightarrow \frac{1}{2},1^{+}\rangle C} &  \\ \mathcal{M}^{JL}_{|A\rangle\rightarrow \frac{3}{2},1^{+}\rangle C}  \end{pmatrix} \\=\begin{pmatrix} \cos\theta &  -\sin\theta  \\ \sin\theta & \cos\theta   \end{pmatrix} \\\begin{pmatrix} \mathcal{M}^{JL}_{|A\rangle\rightarrow ^{3}P_{1}\rangle C} &  \\ \mathcal{M}^{JL}_{|A\rangle\rightarrow ^{1}P_{1}\rangle C}  \end{pmatrix} \\
\end{equation}
and the decay width can be expressed as
\begin{equation}
\begin{split}
& \Gamma(|A\rangle\rightarrow \frac{1}{2},1^{+}\rangle C)=\sum_{JL}|\cos\theta\mathcal{M}^{JL}_{|A\rangle\rightarrow ^{3}P_{1}\rangle C}-\sin\theta\mathcal{M}^{JL}_{|A\rangle\rightarrow ^{1}P_{1}\rangle C}|^{2} \\
& \Gamma(|A\rangle\rightarrow \frac{3}{2},1^{+}\rangle C)=\sum_{JL}|\sin\theta\mathcal{M}^{JL}_{|A\rangle\rightarrow ^{3}P_{1}\rangle C}+\cos\theta\mathcal{M}^{JL}_{|A\rangle\rightarrow ^{1}P_{1}\rangle C}|^{2} \\
\end{split}
\end{equation}

\begin{large}
\textbf{3 Numerical Results}
\end{large}

The input parameters in the $^{3}P_{0}$ model mainly include the
light quark pair($q\overline{q}$) creation strength $\gamma$, the
SHO wave function scale parameter $R$, and the masses of the mesons
and the constituent quarks. The adopted masses of the mesons are
listed in TABLE I, and $m_{u} = m_{d} = 0.22$ GeV, $m_{s} = 0.419$
GeV and $m_{c} = 1.65$ GeV~\cite{Olive}.
\begin{table*}[htbp]
\begin{ruledtabular}\caption{The adopted masses of the mesons used in our calculation.}
\begin{tabular}{c c c c c c c c c c c c c c c c c c c c}
States & \  $M_{\pi^{+}}$  & \ $M_{\pi^{0}}$  & \ $M_{K^{+}}$ &\ $M_{K^{*}}$ & \ $M_{\eta}$ & \ $M_{\eta^{'}}$ & \ $M_{D^{+}}$ & \ $M_{D^{0}}$   \\
\hline
Mass(MeV) & \   139.57         &  \ 134.9766     & \  493.677    &  \   891.66   & \  547.853   &  \  957.78       &  \ 1869.6     &  \  1864.83     \\
\hline
States & \ $M_{D^{*+}_{s}}$  & \ $M_{D^{+}_{s}}$ & \ $M_{D^{*}_{0}(2400)}$  & \ $M_{D(2430)}$ & \ $M_{D(2420)}$  & \ $M_{D^{*\pm}_{s_{0}}(2317)}$ & \ $M_{\rho}$ & \ $M_{\omega}$ \\
\hline
Mass(MeV) & \    2112.3         &  \ 1968.47        & \  2318         &  \   2427        & \  2421.3             &  \ 2317.8           &  \  770   &  \ 782  &  \  \\
\hline
States & \ $M_{D^{*+}}$ & \ $M_{D^{*0}}$ & \ $M_{D^{*}_{2}(2460)}$  & \ $M_{D_{s_{1}}(2460)}$ & \ $M_{D_{s_{1}}(2536)}$ \\
\hline
Mass(MeV) &  \ 2010.25  &  \  2006.96  &  \  2464.4  & \ 2459.5 & \ 2535.11 \\
\end{tabular}
\end{ruledtabular}
\end{table*}
\begin{figure}[h]
\begin{minipage}[t]{0.45\linewidth}
\centering
\includegraphics[height=5cm,width=7cm]{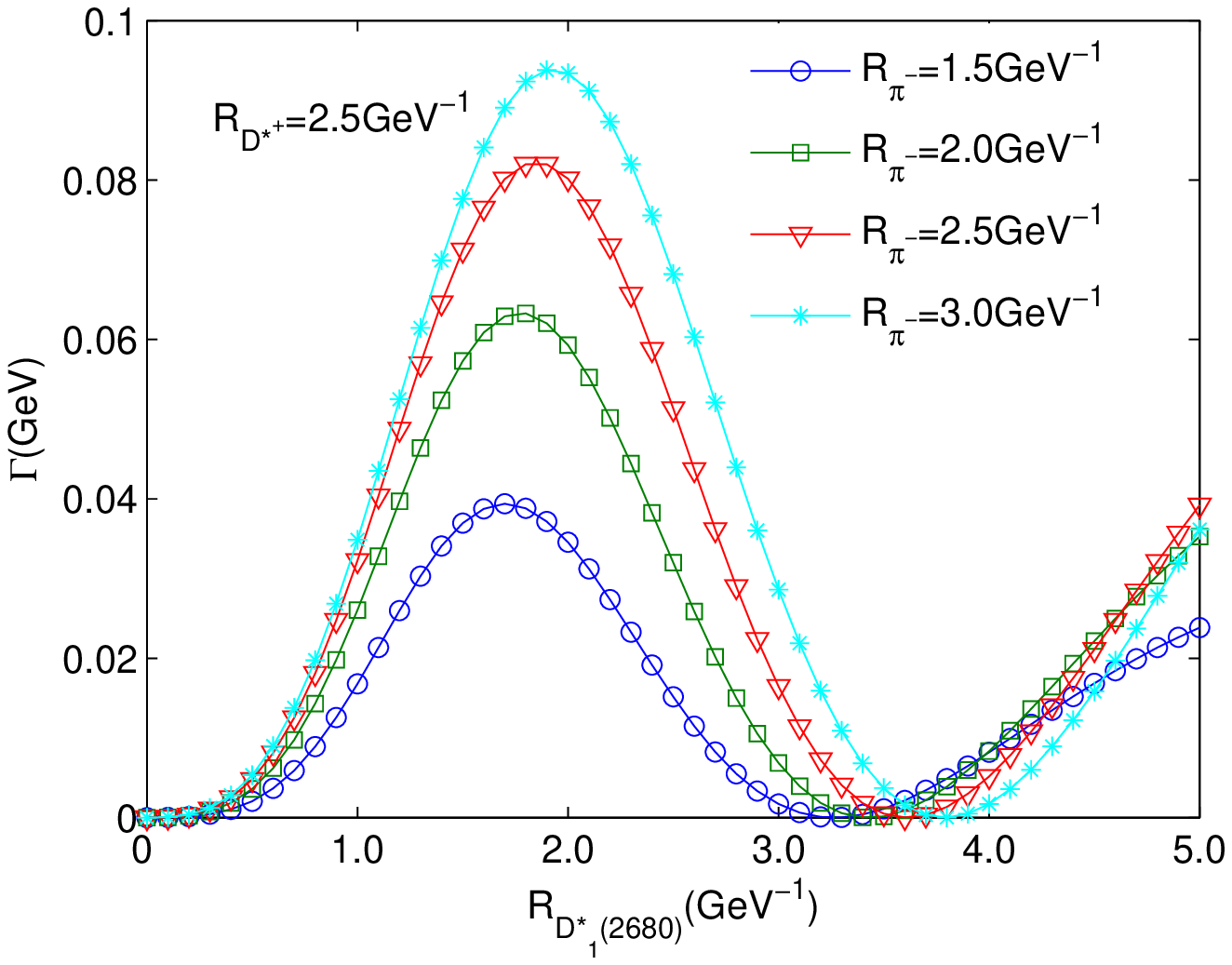}
\caption{The strong decay of $D_{1}^{*}(2680)\rightarrow D^{*+}\pi^{-}$
with $R_{D^{*+}}=2.5$ GeV$^{-1}$.\label{your label}}
\end{minipage}
\hfill
\begin{minipage}[t]{0.45\linewidth}
\centering
\includegraphics[height=5cm,width=7cm]{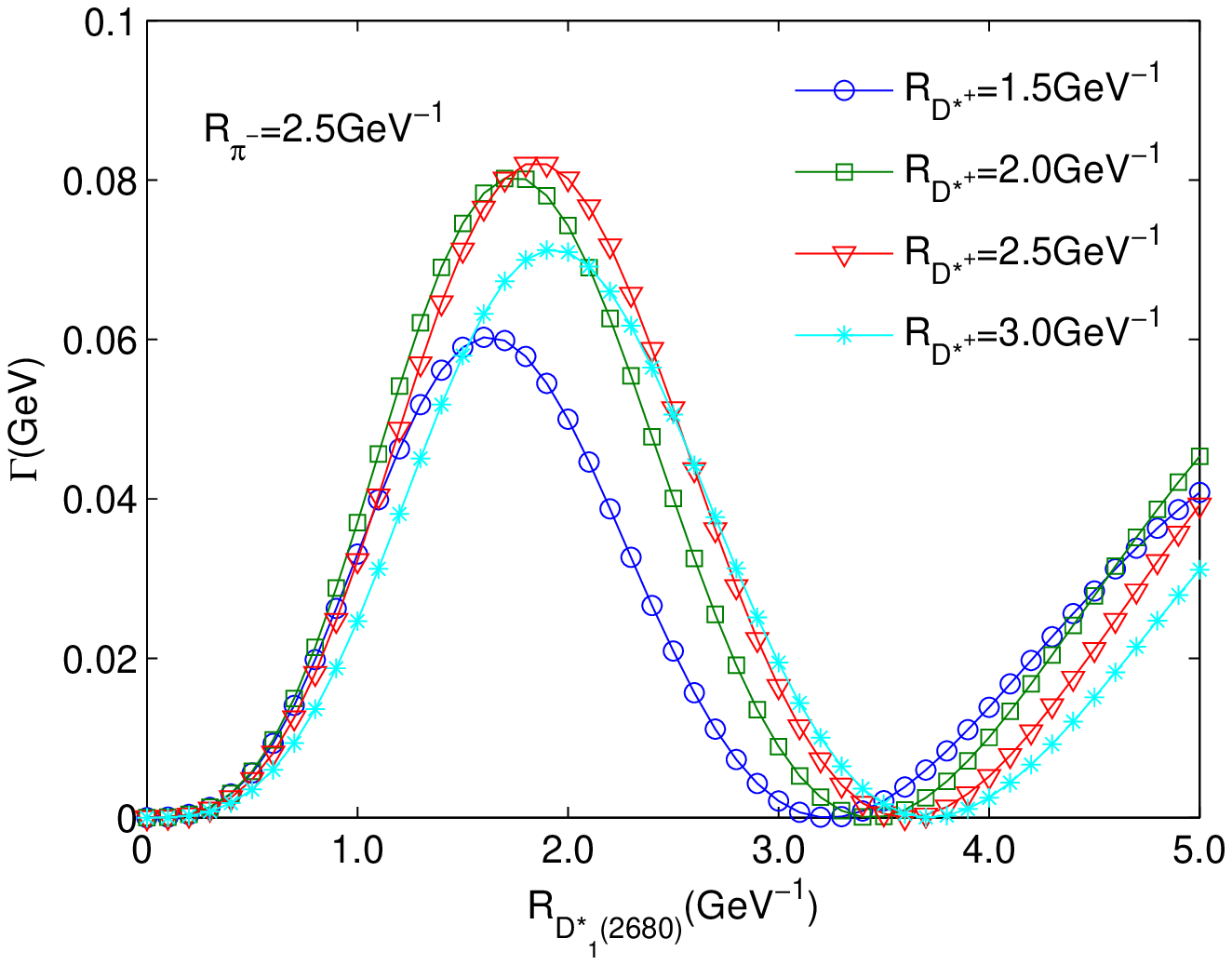}
\caption{The strong decay of $D_{1}^{*}(2680)\rightarrow D^{*+}\pi^{-}$
with $R_{\pi^{-}}=2.5$ GeV$^{-1}$.\label{your label}}
\end{minipage}
\end{figure}

The scale parameter $R$ has a significant influence on the shapes of
the radial wave functions. The spatial integral in Eq.3 is sensitive
to the parameter $R$, therefore the decay width based on the
$^{3}P_{0}$ model is sensitive to the parameter $R$. Taking the
decay $D_{1}^{*}(2680)\rightarrow D^{*+}\pi^{-}$ as an example, we
plot the decay width versus the input parameter $R$ in Figs. 2 and
3. From these two figures, we can easily see the dependence of the
decay width on the input parameter $R$. If $R_{D^{*+}}$ and
$R_{\pi^{-}}$ are all fixed to be $2.5$GeV$^{-1}$(the lines with
triangles in Figs. 2 and 3), the decay width of the
$D_{1}^{*}(2680)$ changes several times with the value of
$R_{D_{1}^{*}(2680)}$ from $1.5$GeV$^{-1}$ to $3.0$GeV$^{-1}$.
Similarly, the decay width changes $2-3$ times, when
$R_{D_{1}^{*}(2680)}$ and $R_{\pi^{-}}$(or $R_{D_{1}^{*}(2680)}$ and
$R_{D^{*+}}$) are fixed to be $2.5$GeV$^{-1}$ while the value of
$R_{D^{*+}}$(or $R_{\pi^{-}}$) changes.

Once the optimal values of $\gamma$ and $R$ are determined, the best
predictions based on $^{3}P_{0}$ decay model are expected. In
reference~\cite{Blun} H.G.Blundel $et$ $al$. carried out a series of
least squares fits of the model predictions to the decay widths of
$28$ of the best known meson decays, and the common oscillator
parameter $R$ with a value of $2.5$GeV$^{-1}$ is suggested to be the
optimal value. As for the factor $\gamma$, it was also fitted at the
same time according to experimental data, giving a fitted value of
$6.25$~\cite{Blun}. More detailed analysis of the input parameters
in the $^{3}P_{0}$ model can be found in Ref.~\cite{Blun}. Thus, we
adopt the SHO wave function with common $R$ whose value is chosen to
be $2.5$GeV$^{-1}$. Correspondingly, the $\gamma$ value is chosen to
be $6.25$ for the creation of $u/d$ quark~\cite{Blun}. As for the
strange quark pair($s\overline{s}$), its creation strength can be
related by $\gamma_{s\overline{s}}=\gamma/\sqrt{3}$~\cite{Yao77}.
\begin{table*}[htbp]
\begin{ruledtabular}\caption{The experimental values and numerical result  based on the  $^{3}P_{0}$ decay model of the ratio $\frac{\Gamma(D_{2}^{*}(2460)\rightarrow D^{+}\pi^{-})}{\Gamma(D_{2}^{*}(2460)\rightarrow D^{*+}\pi^{-})}$ }
\begin{tabular}{c c c c c c c c c c c c c c c c c c}
BaBar~\cite{Amo} & \  CLEO~\cite{Avery94}  & \ CLEO~\cite{Avery90}  & \ ARGUS~\cite{Albr89} &\ ZEUS~\cite{Chek09} & \  3P$_{0}$  \\
\hline
$1.47 \pm 0.03 \pm 0.16$ & \   $2.2 \pm 0.7 \pm 0.6$         &  \ $2.3 \pm 0.8$     & \  $3.0 \pm 1.1 \pm 1.5$    &  \   $2.8\pm0.8^{+0.5}_{-0.6}$   &  \  2.29     \\
\end{tabular}
\end{ruledtabular}
\end{table*}
As a simple test, we also calculate the decay ratio
$\frac{\Gamma(D_{2}^{*}(2460)\rightarrow
D^{+}\pi^{-})}{\Gamma(D_{2}^{*}(2460)\rightarrow D^{*+}\pi^{-})}$ of
$D_{2}^{*}(2460)$ meson with the above parameters.  The
corresponding experimental data from the BaBar ~\cite{Amo},
CLEO~\cite{Avery94,Avery90}, ARGUS~\cite{Albr89}, and
ZEUS~\cite{Chek09} collaborations are listed in TABLE II. The
present calculation $2.29$ based on the $^{3}P_{0}$ model is in
agreement well with the average experimental value 2.35. Certainly,
we can also predict the decay ratio
$\frac{\Gamma(D_{2}^{*}(2460)\rightarrow
D^{+}\pi^{-})}{\Gamma(D_{2}^{*}(2460)\rightarrow D^{*+}\pi^{-})}$
with some other methods such as the heavy-quark symmetry theory~\cite{Rosn}
and the heavy meson effective theory~\cite{WZG2}. With the assumption that the transition is dominated by
$\overline{u}\rightarrow \pi^{-}d$, the heavy-quark symmetry theory gave the expression of the decay ratio
$r=\frac{2}{3}(\frac{p}{p^{*}})^{5}=2.44$,
where $p=507$ MeV and $p^{*}=391$ MeV are the c.m. $3$-momenta in the decays $D_{2}^{*0}\rightarrow
D^{+}\pi^{-}$ and $D_{2}^{*0}\rightarrow D^{*+}\pi^{-}$, respectively.
In reference~\cite{WZG2}, the heavy meson effective theory almost gave the same expression
as it of the heavy-quark symmetry theory.
Thus, our calculation is just a primary verification, which indicates that
the $^{3}P_{0}$ model with the above parameters can reproduce the
experimental data to some extent.

The numerical values of the decay widths and ratios of the charmed
mesons $D_{1}^{*} (2680)$, $D_{3}^{*}(2760)$ and $D_{2}^{*} (3000)$
observed by the LHCb collaboration are presented in TABLE III-IV. It
can be seen from TABLE III that the total width of $D_{3}^{*}(2760)$
is consistent well with the experimental data of LHCb collaboration,
which indicates $D_{3}^{*}(2760)$ is most probably the
$1D\frac{5}{2}3^{-}$ meson. Besides the decay channel
$D^{+}\pi^{-}$, the decay ratios in TABLE IV indicates that the
other probable decay channels include $D^{*+}\pi^{-}$,
$D^{*0}\pi^{0}$, $D_{S}^{+}K^{-}$, $D^{*0}\eta$, $D^{0}\eta$ and
$D^{+}\rho$. As for $D_{1}^{*} (2680)$, the total width is predicted
to be $208.91$MeV which is about $21$ MeV above the central value of
the experimental data. Considering the total uncertainties of the
experimental data, our result is also in agreement with it, which
suggests that $D_{1}^{*} (2680)$ can be assigned as the
$2S\frac{1}{2}1^{-}$ state. Besides $D^{+}\pi^{-}$, $D^{*+}\pi^{-}$,
$D^{0}\pi^{0}$, $D^{*0}\eta$ and $D^{*0}\pi^{0}$ are also its
dominant decay channels.
\begin{table}[htbp]
\begin{ruledtabular}\caption{The strong decay widths of $D_{1}^{*}(2680)$, $D_{3}^{*}(2760)$ and $D^{*}_{2}(3000)$ with possible assignments. If the corresponding decay channel is forbidden,we mark it by "-". All values in units of MeV.}
\begin{tabular}{c c c c c c c }
& \ $D_{1}^{*}(2680)$  & \ $D_{3}^{*}(2760)$  & \multicolumn{4}{c}{$D_{2}^{*}(3000)$}    \\
\hline
       & \ $2S\frac{1}{2}1^{-}$  & \ $1D\frac{5}{2}3^{-}$  & \ $1F\frac{5}{2}2^{+}$ & \ $2P\frac{3}{2}2^{+}$ & \ $2F\frac{5}{2}2^{+}$ & \ $3P\frac{3}{2}2^{+}$  \\
\hline
$D^{*+}\pi^{-}$       &  \  50.92      & \   17.24      &  \   9.67   & \  0.97     &  \   1.45  &  \   3.02     \\
$D_{S}^{*+}K^{-}$     &  \  12.68       & \   0.38     &  \  7.97    & \   24.21    &  \   0.53   &  \ 1.37 \\
$D^{*0}\pi^{0}$       &  \  25.53      & \   8.85     &  \ 4.76     & \  0.43     &  \  0.75   &  \ 1.55  \\
$D^{*0}\eta$          &  \  20.01       & \   13.86      &  \ 8.05    & \  5.52     &  \  0.06  &  \ 0.18    \\
$D^{*0}\eta^{'}$      &  \    -        & \   -      &  \   7.75   & \   16.58    &  \     0.87  &  \ 2.10 \\
$D^{+}\pi^{-}$        &  \  18.17       & \  27.51       &  \  7.17    & \   1.11    &  \   4.85   &  \ 3.86  \\
$D_{S}^{+}K^{-}$      &  \  22.68       & \   2.52      &  \   10.35   & \  11.17     &  \  0.08  &  \ 0.09   \\
$D^{0}\pi^{0}$        &  \   8.86      & \   14.10      &  \  3.46    & \  0.63     &  \  2.47  &  \ 1.96  \\
$D^{0}\eta$           &  \   16.37      & \  5.13       &  \  7.88    & \  0.37     &  \  1.13 &  \  1.04    \\
$D^{0}\eta^{'}$       &  \   -         & \    -     &  \  15.82    & \   9.84    &  \  0.46  &  \ 0.33    \\
$D^{*+}\rho$           &  \  -        & \    -     &  \  15.70   & \  100.10     &  \  0.41 &  \ 7.23    \\
$D^{*+}_{S}K^{*}$      &  \  -      & \      -   &  \  3.27    & \  34.87     &  \   1.09  &  \ 5.74  \\
$D^{*0}\rho$           &  \  -      &  \    -    & \   7.85    & \  50.10     & \    0.19  &  \ 3.51   \\
$D^{*0}\omega$         &  \  -      & \     -    &  \  7.87    & \   50.11    &  \    0.23  &  \ 3.82   \\
$D^{+}\rho$           &  \   15.97   & \   1.18      &  \  17.44    & \  12.51    &  \   0.09  &  \ 0.28   \\
$D_{S}^{+}K^{*}$      &  \   -     & \   -      &  \  8.01   & \  28.72     &  \  1.31 &  \ 3.39   \\
$D^{0}\rho$           &  \  9.22     &  \  0.66      & \   8.63   & \  6.00     & \   0.06  &  \ 0.17    \\
$D^{0}\omega$         &  \  6.28   & \    0.51     &  \  8.82    & \  6.49     &  \   0.04  &  \ 0.12   \\
$D(2420)\pi^{0}$      &  \  2.21   & \   0.01      &  \  5.88    & \  5.13     &  \   0.22  &  \  0.02  \\
$D(2420)\eta$      &  \    -    & \     -    &  \  9.31    & \  1.49     &  \    0.58  &  \ 0.50  \\
$D(2430)\pi^{0}$      &  \  0.01   & \    0     &  \  0.82    & \ 0.79     &  \   0.01  &  \  1.28   \\
$D(2430)\eta$      &  \   -     & \     -    &  \  1.49    & \  1.99     &  \   1.32  &  \ 0.69  \\
$D^{*}_{0}(2400)\pi^{0}$  &  \   -     & \     -    &  \  0   & \   0    &  \  0   &  \ 0 \\
$D^{*}_{0}(2400)\eta$  &  \   -    & \    -     &  \   0   & \   0    &  \  0  &  \ 0  \\
$D_{S}(2460)K^{-}$     &  \   -     & \   -     &  \  1.61    & \  3.45    &  \   0.52 &  \  0.34  \\
$D_{S}(2536)K^{-}$     &  \   -    & \    -     &  \  10.14    & \  1.39    &  \  2.78  &  \ 1.47   \\
$D_{2}^{*+}(2460)\pi^{-}$  &  \   -     & \   0.65      &  \ 16.73    & \  39.69     &  \    5.93 &  \ 10.46   \\
$D_{2}^{*0}(2460)\pi^{0}$  &  \   -     & \    0.32     &  \ 8.38     & \  19.88     &  \   2.97 &  \ 5.24   \\
$D_{2}^{*0}(2460)\eta$  &  \    -    & \    4.49     &  \  5.22    & \   11.82    &  \    1.69 &  \ 2.81   \\
$D_{s_{0}}^{*+}(2317)K^{-}$  &  \  -     & \   -      &  \  0    & \  0     &  \  0  &  \  0   \\
Total width           &  \   208.91    & \    97.41     &  \  220.05    & \  442.36     &  \   32.09 &  \  62.57   \\
\end{tabular}
\end{ruledtabular}
\end{table}
\begin{table}[htbp]
\begin{ruledtabular}\caption{The decay ratios of partial decay width $\Gamma_{p}/\Gamma_{T}$ of $D_{1}^{*}(2680)$, $D_{3}^{*}(2760)$ and $D^{*}_{2}(3000)$ with possible assignments.}
\begin{tabular}{c c c c c c c}
& \ $D_{1}^{*}(2680)$  & \ $D_{3}^{*}(2760)$  & \multicolumn{4}{c}{$D_{2}^{*}(3000)$}   \\
\hline
       & \ $2S\frac{1}{2}1^{-}$  & \ $1D\frac{5}{2}3^{-}$  & \ $1F\frac{5}{2}2^{+}$ & \ $2P\frac{3}{2}2^{+}$ & \ $2F\frac{5}{2}2^{+}$   &  \ $3P\frac{3}{2}2^{+}$ \\
\hline
$D^{*+}\pi^{-}$       &  \  0.24      & \   0.18      &  \   0.04   & \  0.002     &  \   0.05  &  \ 0.05    \\
$D_{S}^{*+}K^{-}$     &  \  0.06       & \   0.004     &  \  0.04    & \   0.05    &  \   0.02 &  \ 0.02   \\
$D^{*0}\pi^{0}$       &  \  0.12      & \   0.09     &  \ 0.02     & \  0.001     &  \  0.02 &  \ 0.02  \\
$D^{*0}\eta$          &  \  0.10       & \   0.14      &  \ 0.04    & \  0.01     &  \  0.002 &  \  0.003    \\
$D^{*0}\eta^{'}$      &  \    -        & \   -      &  \   0.04   & \   0.04    &  \     0.03  &  \ 0.03 \\
$D^{+}\pi^{-}$        &  \  0.09       & \  0.28       &  \  0.03    & \   0.003    &  \   0.15 &  \ 0.06    \\
$D_{S}^{+}K^{-}$      &  \  0.11       & \   0.03      &  \   0.05   & \  0.03     &  \  0.003 &  \ 0.001    \\
$D^{0}\pi^{0}$        &  \   0.04      & \   0.14      &  \  0.02    & \  0.001     &  \  0.08 &  \ 0.03   \\
$D^{0}\eta$           &  \   0.08      & \  0.05       &  \  0.04    & \  0.0008     &  \  0.04 &  \ 0.02     \\
$D^{0}\eta^{'}$       &  \   -         & \    -     &  \  0.07    & \   0.02    &  \  0.01 &  \  0.005    \\
$D^{*+}\rho$           &  \  -        & \    -     &  \  0.07   & \  0.23     &  \  0.01 &  \ 0.12    \\
$D^{*+}_{S}K^{*}$      &  \  -      & \      -   &  \  0.01   & \  0.08     &  \   0.03  &  \ 0.09  \\
$D^{*0}\rho$           &  \  -      &  \    -    & \   0.04    & \  0.11     & \    0.006 &  \  0.06   \\
$D^{*0}\omega$         &  \  -      & \     -    &  \  0.04    & \   0.11    &  \    0.007 &  \  0.06   \\
$D^{+}\rho$           &  \   0.08   & \   0.01      &  \  0.08    & \  0.03    &  \   0.003 &  \ 0.004    \\
$D_{S}^{+}K^{*}$      &  \   -     & \   -      &  \  0.04   & \  0.06     &  \  0.04 &  \ 0.05   \\
$D^{0}\rho$           &  \  0.04     &  \  0.007      & \   0.04   & \  0.01     & \   0.002 &  \  0.003    \\
$D^{0}\omega$         &  \  0.03   & \    0.005     &  \  0.04    & \  0.01     &  \   0.001  &  \ 0.002   \\
$D(2420)\pi^{0}$      &  \  0.01   & \   0.0001      &  \  0.03    & \  0.01     &  \   0.007 &  \  0.0003   \\
$D(2420)\eta$      &  \    -    & \     -    &  \  0.04    & \  0.003     &  \    0.02 &  \  0.008  \\
$D(2430)\pi^{0}$      &  \  0   & \    0     &  \  0.004    & \ 0.002     &  \   0.0003 &  \  0.02    \\
$D(2430)\eta$      &  \   -     & \     -    &  \  0.007    & \  0.005     &  \   0.04 &  \ 0.01   \\
$D^{*}_{0}(2400)\pi^{0}$  &  \   -     & \     -    &  \  0   & \   0    &  \  0  &  \  0 \\
$D^{*}_{0}(2400)\eta$  &  \   -    & \    -     &  \   0   & \   0    &  \  0 &  \  0  \\
$D_{S}(2460)K^{-}$     &  \   -     & \   -     &  \  0.007    & \  0.008    &  \   0.02 &  \ 0.005   \\
$D_{S}(2536)K^{-}$     &  \   -    & \    -     &  \  0.05    & \  0.003    &  \  0.09 &  \  0.02   \\
$D_{2}^{*+}(2460)\pi^{-}$  &  \   -     & \   0.007      &  \ 0.08    & \  0.08     &  \    0.18 &  \ 0.17   \\
$D_{2}^{*0}(2460)\pi^{0}$  &  \   -     & \    0.003     &  \ 0.04     & \  0.04     &  \   0.09 &  \  0.08  \\
$D_{2}^{*0}(2460)\eta$  &  \    -    & \    0.05     &  \  0.02    & \   0.03    &  \    0.05  &  \  0.04 \\
$D_{s_{0}}^{*+}(2317)K^{-}$  &  \  -     & \   -      &  \  0    & \  0     &  \  0   &  \  0  \\
\end{tabular}
\end{ruledtabular}
\end{table}

Experiments indicate $D^{*}_{2}(3000)$ is a $2^{+}$ state charmed
meson~\cite{Aaij}. Thus, we study its decay behavior with the
$1F\frac{5}{2}2^{+}$, $2P\frac{3}{2}2^{+}$, $2F\frac{5}{2}2^{+}$ and
$3P\frac{3}{2}2^{+}$ assignments. As the candidate of
$D^{*}_{2}(3000)$, the total width of $2F\frac{5}{2}2^{+}$ is
predicted to be only $32.09$MeV which is about $150$MeV smaller than
the central value of the experimental data. Thus, it can be
completely excluded from the probable assignments. In addition, it
can be seen from TABLE III that the width of $2P\frac{3}{2}2^{+}$ is
about $120$MeV above the upper limit of the experimental data. Thus,
$D^{*}_{2}(3000)$ is also impossible to be the $2P\frac{3}{2}2^{+}$
state. In addition, if $D^{*}_{2}(3000)$ is $3P\frac{3}{2}2^{+}$
state, its predicted cross section is $62.57$MeV which is smaller
about $123$MeV than the central value of the experimental data.
Although the calculated total width is just above the lower limit of
the experimental data, its branching ratio of the $D^{+}\pi^{-}$
decay channel is very small. Thus, $3P\frac{3}{2}2^{+}$ state is
also less likely to be the assignment of $D^{*}_{2}(3000)$.

Although the predicted value of the total width of
$1F\frac{5}{2}2^{+}$ is somewhat bigger than the central value of
experimental data, it is within the error range. This indicates
$1F\frac{5}{2}2^{+}$ is most likely to be the assignment of
$D^{*}_{2}(3000)$. However, this determination needs to be further
verified according to experiments in the future. We can see from
Table IV that no decay channel show an obvious advantage over
another, while the $D^{*}_{2}(3000)$ resonance is observed by the
LHCb Collaboration in the  $D^{+}\pi^{-}$ channel. One possible
explanation about this behavior is that the production cross section
of $D^{*}_{2}(3000)$ is so large that the fairly small branching
ratio is still observable. If the decay ratios of different decay
channels are measured in experiments in the future, this
determination can be exactly verified. At present, we can
temporarily assign $D^{*}_{2}(3000)$ charmed meson as the
$1F\frac{5}{2}2^{+}$ state, while $2F\frac{5}{2}2^{+}$,
$2P\frac{3}{2}2^{+}$ and $3P\frac{3}{2}2^{+}$ states can be excluded
temporarily. Nevertheless, these decay predictions for the
$2F\frac{5}{2}2^{+}$, $2P\frac{3}{2}2^{+}$ and $3P\frac{3}{2}2^{+}$
states are valuable in further searches for the partners of
$D^{*}_{2}(3000)$. For $2P\frac{3}{2}2^{+}$ as an example, its decay
ratios of $D^{*+}\rho$, $D^{*0}\rho$ and $D^{*0}\omega$ is much more
obvious than the other decay modes, which can be used as a valuable
judgement of this meson.

In reference~\cite{JZW}, the decay behavior of $D^{*}_{2}(3000)$ was
also analyzed using the $^{3}P_{0}$ decay model. The
$3P\frac{3}{2}2^{+}$ state was predicted as the most possible
assignment of the $D^{*}_{2}(3000)$ in their work, while the
assignment of the $2F\frac{5}{2}2^{+}$ charmed meson could not be
fully excluded. The primary difference between our analysis and theirs
in reference~\cite{JZW} about the $D^{*}_{2}(3000)$ charmed meson is that they employed the SHO
wave function with the effective scale parameter $R$~\cite{JZW},
while we adopt the common valve of the scale parameter $R$ which was
calculated by fitting the experimental data in
reference~\cite{Blun}. Thus, the difference between the results in
reference~\cite{JZW} and ours is mainly due to the influence of the
input parameter $R$, which needs further confirmation by future
experimental data from LHCb and forthcoming Belle II.

\begin{large}
\textbf{4 Conclusion}
\end{large}

In this article, we carry out an analysis of the newly observed
charmed mesons  $D_{1}^{*}(2680)$, $D_{3}^{*}(2760)$ and
$D_{2}^{*}(3000)$ reported by LHCb collaboration with the
$^{3}P_{0}$ decay model. Our analysis supports $D_{1}^{*}(2680)$ and
$D_{3}^{*}(2760)$ to be the $2S\frac{1}{2}1^{-}$ and
$1D\frac{5}{2}3^{+}$ assignments separately. In addition, the
partial width and ratios are obtained, further shedding light on the
nature of these two mesons. The total width predicted by the
$^{3}P_{0}$ decay model supports the $1F\frac{5}{2}2^{+}$ for the
$D_{2}^{*}(3000)$ meson, which needs further confirmation from the
measured partial decay ratios. When investigating $D_{2}^{*}(3000)$,
we have also analyzed the decay behavior of the $2P\frac{3}{2}2^{+}$
, $2F\frac{5}{2}2^{+}$ and $2P\frac{3}{2}2^{+}$ states, which can be
used as valuable judgements for the assignments of the newly
observed charmed mesons in the future.

\begin{large}
\textbf{Acknowledgment}
\end{large}

This work is supported by National Natural Science Foundation of
China, Grant Numbers 11375063 and the Fundamental Research Funds for
the Central Universities, Grant Number 2016MS133 and 13QN59.

\end{document}